# Analysis of temporal structures of seismic events on different scale levels

V. I. German

Siberian Aerospace State University named after M. F. Reshetnev, Krasnoyarsk, Russia

Correspondence to: V. I. German (germanv@rambler.ru)

**Abstract**

A statistical model for describing the scaling of the distribution of inter-event times is described. By considering the diverse region seismicity (natural and induced) at different scale levels the self-similarity of the distribution has been determined. Significant deviations occur only in the area of maximum magnitude (maximum energy class). A comparison between the distribution of inter-event times at different scale levels and the most popular distributions of reliability theory has been carried out. The distribution of inter-event times for different scale levels is well approximated by the Weibull distribution. The Weibull distribution, with parameters which obey the scaling model and the Gutenberg-Richter law, have been tested.

## 1  Introduction

The temporal structure of a seismic process, together with the spatial and energy structures are general characteristics required for its analysis and for forecasting strong seismic events. For a fixed space area, the temporal structure is determined by the set of the inter-event times for each scale level of seismicity (IETSL).

The average value of IETSL is inversely proportional to the number of events for each level. In this sense, they have been used for a long time to construct Gutenberg-Richter (G-R) plots, which show the self-similarity in energy radiation for different scale levels. At the same time, the hierarchy in energy radiation in time has been found (Rykunov et al., 1987). These relationships confirm the self-similarity of seismic/failure process in the earth's crust/rocks on different scale levels: from micro-level to strong earthquakes (Sadovskiy et al., 1987;

Rykunov et al., 1987; Turcotte 1992; Kuksenko et al., 1996; Mukhamedov, 1996; Ulomov, 1998; Bak at al., 2002; Corral, 2003; Corral, 2004, Baiesi and Paczuski, 2004).

The self-similarity of the seismic process at different scale levels allows the strong seismic events to be forecasted by analyzing the weak seismicity. For example, the G-R plot constructed for short time measurements (with a large number of weak events and maybe without any strong ones) can be used to estimate the number of events for all levels of seismicity, at least up to the magnitude $M_{LH} = 6$ (Ulomov et al., 1999). Moreover, the recognition of the deviation of current temporal characteristics from long term ones has been used for intermediate and short-term forecasting of strong events (Sobolev et al., 1991; Knopoff et al., 1996). The inter-event times or seismic activities (without division into scale levels) have been used for the same purpose (e.g. Tomilin and Voinov, 1995).

Detailed investigation of the IETSL distribution allows additional information about the regularities of seismic process to be obtained and to be used for forecasting strong seismic events forecasting.

## 2  The properties of inter-event times for different scale levels of seismicity

The conception of the self-similarity of seismic process on different scale levels includes also a temporal self-similarity. Let's consider a case, which corresponded to the self-similarity of inter-event times in a general case: that they have the same type of distribution but have a scale coefficient $\varphi(\vec{\varepsilon})$. This is a function of the energy-time-area interval considered, which is defined by the vector of its parameters $\vec{\varepsilon}$. $\varphi(\vec{\varepsilon})$ should take into account the seismic activity variation for different energy-time-area intervals.

This corresponds to the statistical accelerated life model (Cox and Oakes, 1984), which proposed that the lifetime of object at a scale level is obtained as a pressed or unpressed version of the life time of the object, on base level $\vec{\varepsilon} = \vec{\varepsilon}_0$.

Thus, according to this model the IETSL distribution function $F(\Delta t, \vec{\varepsilon})$ will depend on a temporal variable, $\Delta t$, and on an additional (scale) vector variable $\vec{\varepsilon}$, and will obey the relationship:

$$F(\Delta t, \vec{\varepsilon}) = F[\Delta t \varphi(\vec{\varepsilon}), \vec{\varepsilon}_0] = F_0(\Delta t \varphi(\vec{\varepsilon}))$$

and the density function, $f(\Delta t, \vec{\varepsilon})$, will obey:

$$f(\Delta t, \vec{\varepsilon}) = \varphi(\vec{\varepsilon}) f[\Delta t \varphi(\vec{\varepsilon}), \vec{\varepsilon}_0] = \varphi(\vec{\varepsilon}) f_0(\Delta t \varphi(\vec{\varepsilon})).$$

The variable $\Delta t$ corresponds to the stochastic variable $\Delta T$, which is the inter-event times for the energy-time-area interval, defined by the vector of its parameters $\vec{\varepsilon}$. The relationship for $F(\Delta t, \vec{\varepsilon})$ allows the following equation for $\Delta T$ to be written:

$$\Delta T = \Delta T_\varepsilon = \Delta T_{\varepsilon_0} / \varphi(\vec{\varepsilon}). \tag{1}$$

To determine the scale parameter of the model, it is necessary to calculate a mean for both parts of Eq. (1), find the logarithm of them, multiply to $-1$ and then to carry out the substitution $-\log_{10}[M(\Delta T_\varepsilon)] \cong -\log_{10}(\sum_i \Delta t_{\varepsilon_i} / N_\varepsilon) = \log_{10}(N_\varepsilon / T)$:

$$\log_{10}(N_\varepsilon / T) = \log_{10}[\varphi(\vec{\varepsilon})] - \log_{10}[M(\Delta T_{\varepsilon_0})], \tag{2}$$

where $T$ is the duration of the considered time period and $N_\varepsilon$ is the number of events for the energy-time-area interval considered. Thus, the general form for the scale coefficient is:

$$\varphi(\vec{\varepsilon}) = M(\Delta T_{\varepsilon_0})(N_\varepsilon / T). \tag{3}$$

Thus it is proportional to seismic activity in the energy-time-area interval considered ($M(\Delta T_{\varepsilon_0}) = const$). Such a scale coefficient was successfully applied by Corral (Corral, 2004) for inter-event times scaling. A comparison with the generalized G-R law:

$$\log_{10}(N_\varepsilon / T) = -\gamma \log_{10} E + d_L \log_{10} L + \log_{10} d_\varepsilon,$$

where $E$ is the radiated energy of seismic event, $L$ characterizes the size of the considered area and $\gamma$, $d_L$ are constants and $d_\varepsilon$ is a average seismic activity when $-\gamma \log_{10} E + d_L \log_{10} L = 0$. In this case, $N_\varepsilon / T = d_\varepsilon E^{-\gamma} L^{d_L}$ and the scale parameter can be written as $\varphi(\vec{\varepsilon}) = (d_\varepsilon M(\Delta T_{\varepsilon_0})) E^{-\gamma} L^{d_L}$; a similar scaling was used in (Corral 2003). In this case, the G-R law will be only a corollary of the inter-event times distribution function self-similarity for different energy-time-area intervals.

To obtain a model for inter-event times for the different energy-time-area intervals, it is necessary to find the logarithm of Eq. (1) and to carry out the substitution $\log_{10} \Delta T_{\varepsilon_0} = M(\log_{10} \Delta T_{\varepsilon_0}) + \theta$:

$$\log_{10} \Delta T = \rho_0 - \log_{10} \varphi(\vec{\varepsilon}) + \theta, \tag{4}$$

where $\rho_0 = M(\log_{10} \Delta T_{\varepsilon_0})$, $\theta$ is the stochastic variable with a mean equal to zero and a distribution function which does not depend on $\vec{\varepsilon}$ (Cox and Oakes, 1984). According to this model, the distribution of $\log_{10} \Delta T$ for different values of $\vec{\varepsilon}$ are the same, but has shifted, thus the standard deviation is constant for all energy-time-area intervals:

$$\sigma(\log_{10} \Delta T) = const. \tag{5}$$

The mean for both parts of Eq. (3):

$$M(\log_{10} \Delta T) = -\log_{10} \varphi(\vec{\varepsilon}) + \rho_0. \tag{6}$$

Equations (5) and (6) are valid if the general assumption about the united scaling law for inter-event times is correct. However, it seems impossible to avoid a significant deviation for all values of inter-event times (Bak et al., 2002; Corral, 2004).

Thus, in this paper, to obtain solid results about self-similarity of IETSL, only the energy scaling has been considered. To calibrate the accelerated life model for each region, the classical G-R law:

$$\log_{10}(N_E / T) = -\gamma \log_{10} E + \log_{10} d_E$$

has been used, where $N_E$ is the number of events, with the logarithm of radiated energy given by: ($\log_{10} E(J) - 0.5; \log_{10} E(J) + 0.5$).

In this case, $\varphi(\vec{\varepsilon}) = \varphi(E) = d / E^{\gamma}$ where $d = d_E M(\Delta T_{\varepsilon_0}) = const$ and

$$M(\log_{10} \Delta T) = \gamma \log_{10} E + \delta_0, \tag{7}$$

where $\delta_0 = \rho_0 - \log_{10} d$ and $\Delta T$ is IETSL for each region. Thus the mean of IETSL logarithm $M(\log_{10} \Delta T)$ has a linear relationship with the energy class $\log_{10} E(J)$. The constant value of standard deviation $\sigma(\log_{10} \Delta T)$ allows the classical form of the least squares method to be utilized to estimate the parameters $\gamma$ and $\delta_0$ effectively. This is an advantage of this model over the G-R law. Another is that, according to the accelerated life model, it is possible to not only estimate the average number of strong seismic events, but also the distribution of inter-event times for them, by analyzing the temporal structure of the weak events. It is also possible to apply the accelerated life model for scaling the distribution of other physical values, such as, for example, the distances between seismic events.

## 3 Experimental data

Experimental checking of the inferences concerning the properties of IETSL structure has been carried out for two types of seismic events, induced and natural, over a wide range of scales of seismic process, covering ~14 orders of magnitude of radiated energy or 14 energy classes (Table 1). For the analysis of natural seismicity, the *Special Catalogue of Earthquakes of Northern Eurasia* (to calculate the energy class, the Rautian formula $\log_{10} E(J) = 1.8 M_{LH} + 4.0$ was used) (Special…) and a catalogue of earthquakes in the Toktogul region ( $M = (\log_{10} E(J) - 1.3)/3$ for $M < 1.8$, $M = (\log_{10} E(J) - 4)/1.8$ for $M \geq 1.8$ ) (Database…), created at the United Institute of the Physics of the Earth of the Russian Academy of Sciences have been used. The catalogue (Special…) does not include foreshocks and aftershocks, which have been removed using the methods described in (Molchan and Dmitrieva, 1992).

Induced seismicity is represented by catalogues of the North Ural Bauxite Mine and Goldfields Welkome (ISS International, South Africa). The average seismic activity for this catalogues does not have a great variation in the period considered. The daily variability of the mining activity on North Ural Bauxite Mine influenced the inter-event times distribution up to energy class $\log_{10} E(J) = 3.5$. For the North Ural Bauxite Mine (German, 2002; German and Mansurov, 2002; Tomilin and Voinov, 1995), only the local area, with high activity and large number of strong rockbursts was considered. At the same time, the large areas were considered for earthquakes (Table 1).

In each region, the G-R plots were used to determine the energy class completeness threshold, $\log_{10} E_{\min}(J)$, above which a catalogue can be considered to be reasonably complete. This corresponds to the point where the G-R plot has a significant change of its slope. The large number of data in the catalogues allows a detailed analysis of the IETSL structure to be carried out. Further important relationships are demonstrated with examples from the North Ural Bauxite Mine and the Kamchatka-Kurils seismoactive region, as extreme scale level representatives.

## 4 Checking of inter-event times self-similarity on different scale levels

The experimental relationships (5) and (7) obtained for the mean and for the standard deviation of $\log_{10} \Delta T$ is shown in Fig. 1. Each point on this figure was obtained for seismic events in the energy interval $(E/10^{0.5}; 10^{0.5} E)$; that is, from one energy class $(\log_{10} E(J) - 0.5; \log_{10} E(J) + 0.5)$. This interval size allows a sufficient number of data to be picked out for the analysis and to control the local features of IETSL structure. A step of the shift for this interval is 0.1 for North Ural Bauxite Mine and 0.2 for all other regions. The same division into scale levels will be used for the further analysis.

Intervals corresponding to the accelerated life model (i.e. they obey equations (5) and (7)) exist for all considered regions; they begin at the energy class completeness threshold $\log_{10} E_{\min}(J)$ (see Table 1). The linear part of the $M(\log_{10} \Delta T)$ plot continues to nearly the maximum energy classes, except at the North Ural Bauxite Mine, where the strongest rockbursts, with $\log_{10} E(J) > 6.5$, do not have clearly induced nature, but mixed induced-tectonic nature (Tomilin and Voinov, 1995). This feature leads to a bimodal (two linear intervals) type of G-R plot. At the same time, $\sigma(\log_{10} \Delta T)$ in areas of strong events often has a significant deviation from a constant value. The existence of intervals which obey the accelerated life model for all regions, allows the parameters $\gamma$ and $\mu_0$ to be used as seismic regime characteristics for each of regions like the parameters of the G-R law are used. The relationships (5) and (7) can be also used to determine the energy class completeness threshold.

The behavior of the curves in Fig. 1a, for the interval $\log_{10} E(J) \in (3.3; 5.7)$, corresponds to the model described and hence the seismic process at different scale levels for this interval is self-similar. This energy interval correlates well with estimates obtained by German and Mansurov (German and Mansurov, 2002). A linear least squares approximation of the relationship is $M(\log_{10} \Delta T(days)) = 0.68 \log_{10} E(J) - 2.24$. The same situation for the interval $\log_{10} E(J) \in (13.0; 16.7)$ is given in Fig. 1b: $M(\log_{10} \Delta T(days)) = 0.45 \log_{10} E(J) - 5.66$.

## 5  Investigation of inter-event times distribution type

The next step in the IETSL investigation was to determine its distribution. Essentially a distribution must be determined which describes all the experimental data, obeys both the G-R law and the accelerated life model and further allows the interaction between events to be described; in other words (mathematically) aftereffect for events arising. The last point is very important, because the interaction leads the seismic process from one level to the next, higher one.

In many papers, the question about inter-event times distribution is considered, but in most cases without division into different scale levels and thus without checking its self-similarity properties. The most popular distribution for inter-event times is the exponential distribution. This corresponds to a simple stream (stationary Poisson process), but in most cases the approximation of the experimental data to an exponential distribution is not sufficiently good. At the same time, some researchers have shown that for some data sets it is possible to obtain a good fit by applying the lognormal, gamma or Weibull distribution (e.g. Rikitake, 1976; Nishenko and Buland, 1987; Mukhamedov, 1996; Correig at al., 1997; Corral, 2004), which have the similar shape in this case.

To estimate the type of IETSL distribution, the dimensionless relationship between the normalized third central moment, $\mu_3 = M[(\log_{10} \Delta T)^3]/\sigma^3(\log_{10} \Delta T)$, and the coefficient of variation, $\chi = \sigma(\log_{10} \Delta T)/M(\log_{10} \Delta T)$, for all regions was constructed for scale levels larger than the energy class completeness threshold, $\log_{10} E_{\min}(J)$ (Table 1). These were then compared with the theoretical curves for the most popular families of distributions of the reliability theory (Cox and Oakes, 1984). In such a coordinate system, each of these families of distributions, for all values of the parameters, is represented as a single curve. This allows a very quick and easy comparison of the experimental data to be made with a large number of families of distributions simultaneously and to obtain an initial estimation of their distribution type. This method should be considered as a variant of a statistical method of moments and hence not as a precise method.

It was determined that the experimental points for all regions are closer to the curve which corresponds to the Weibull family of distributions and near to the curve of the gamma family of distributions (both these families include the exponential family of distributions) (Fig. 2).

The experimental points in Fig. 2 and in figures for other regions can be divided into two groups: the minimum deviation from the theoretical curve is for the energy intervals, which corresponded to weak events, with a coefficient of variation generally >1.3. The other points correspond to the intervals with relatively strong events; for these points, the coefficient of variation is near 1, as it is for an exponential distribution. The coefficient of variation for different scale levels is interesting because it characterizes the clustering of events (Kagan and Jackson, 1991); $\chi = 1$ corresponds to random events, and $\chi > 1$ to clustered ones. Thus even in the catalogue (Special…), with removed aftershocks, weak events are clustered.

An advantage of the Weibull distribution over the exponential one is that an additional attenuation parameter, responsible for aftereffects or clustering; this allows the nonstationary nature of seismic processes to be taken into account. Moreover, application of the Weibull distribution has some theoretical considerations: it is the limiting distribution ($n \to \infty$) for the least variable, from $n$ independent stochastic variables with the same distribution (Gnedenko, 1962). Some models generate the Weibull distribution; for example Mukhamedov (Mukhamedov, 1996) argues that the relaxation processes in a discrete hierarchy medium like rock (the Earth's crust) obeys the Weibull distribution. The Benioff-Shimazaki model, which describes stress release drop (Lomnitz, 1994), also generates it. An additional argument for the Weibull distribution is its successful application in reliability theory.

To test the assumption about the type of IETSL distribution, the Kolmogorov test for composite hypotheses (Lemeshko and Postovalov, 2001), with the maximum-likelihood estimation of distribution parameters, was used. The maximum deviation of the empirical distribution function from the theoretical one $\Delta F_{\max} = (6\sqrt{n}S_c - 1)/(6n)$, where $n$ is the number of elements of the analyzed sample and $S_c$ is a value which is defined by the type of tested distribution and the confidence level of the test (Lemeshko and Postovalov, 2001). It is evident that the test becomes much more restrictive with increasing $n$: if the strongest considered scale level in a region corresponds to the energy class $\log_{10} E_c(J)$ with $n \approx 10$ then $\Delta F_{\max} \approx 0.32 S_c$, due to the G-R law for a level with $\log_{10} E_c(J) - 2$ the number of events is about 100 and $\Delta F_{\max} \approx 0.10 S_c$ and for $\log_{10} E_c(J) - 4$ $n \approx 1000$ and $\Delta F_{\max}$ is only $0.03 S_c$ (for the most popular cases $S_c$ is near to 1). Note that Kolmogorov test for composite hypotheses (when the parameters of distribution are determined from the same experimental data) is more powerful than $\chi^2$-type tests.

The empirical distribution function of IETSL for different values $E$ has enough small deviation from Weibull distribution function ($F(\Delta t) = 1 - \exp(-\lambda \Delta t^k)$, hazard function or failure rate function $r(\Delta t) = \lambda k \Delta t^{k-1}$) not to reject the hypothesis about the Weibull distribution of IETSL with confidence level 0.001 generally for all energy classes, which are large then energy class completeness threshold $\log_{10} E_{\min}(J)$ for the region considered (Table 2). With the same confidence level, the hypothesis for the gamma distribution of IETSL is rejected nearly as often as the Weibull distribution and hypothesis for lognormal distribution is not rejected for strong events. Testing of the hypothesis for the special case of the Weibull distribution – exponential distribution: $F(\Delta t) = 1 - \exp(-\lambda_0 \Delta t)$, $r(\Delta t) = \lambda_0$ shows that it is not rejected only for relatively strong events (Table 2). These results confirm the conclusion of the analysis shown in Fig. 2.

The Weibull distribution corresponds to the Weibull process, which is restarted after each event (Bogdanoff and Kozin, 1985) and which can be obtained from a simple stream by addition the aftereffects. According to the expression for the Weibull intensity function for $k < 1$ ($k$ is shape parameter corresponding to the aftereffect), the probability of occurrence of the next event at a constant time interval decreases with time; thus events have a tendency to group (see also Rykunov et al., 1987). Thus, the results obtained correspond to the idea concerning event interactions and thus to time grouping of weak events, which leads to the occurrence of strong events (e.g. Kuksenko et al., 1996). However, the strongest events are not included in the next level events preparation process and have no interaction; this situation corresponds to $k = 1$ or an exponential distribution with $\chi = 1$ (without aftereffect or without clustering). At the same time, it is extremely difficult to find any physical interpretation for the gamma distribution. For these reasons, the Weibull distribution was preferred.

Fig. 3 shows the behavior of the Weibull distribution parameters $k$, $\lambda$ and, for comparison, the intensity of the exponential distribution $\lambda_0$ versus scale level value. $\lambda$ and $\lambda_0$ are scale parameters which determine the intensity of events occurrence. The shape parameter $k$ corresponding to the aftereffect value is nearly a constant value and less then 1. It has a gradual decreasing trend and a local increase in the area of the strongest events. At the same time, the scale parameter $\lambda$ decreases monotonically as the energy class grows; that is, the activity of the strong seismic events is less than that of the weak ones.

# 6 Relationships between inter-event times properties and the Weibull distribution parameters

The next step for IETSL investigation is to establish the restrictions on the Weibull distribution parameters to obey the accelerated life model and the G-R law. For the stochastic variable $\Delta T$, which corresponds to IETSL and has a Weibull distribution:

$$M(\log_{10} \Delta T) = -\log_{10}(\lambda e^C)/k, \qquad (8)$$

where $C \approx 0.577$ is the Euler number and

$$\sigma(\log_{10} \Delta T) = \pi \log_{10} e/(6^{0.5} k) \approx 0.557/k. \qquad (9)$$

According to equations (8) and (9), to satisfy the conditions of the accelerated life model (5), (7), it is necessary to fix $k$, in order to have a constant value for the standard deviation of the IETSL logarithm and to have a linear relationship between $\log_{10} \lambda$ and the energy class. Therefore, the scale variable for seismic process is the parameter $\lambda$. The same result can be obtained by comparing the general equation for cumulative distribution scaling and the expression for the Weibull distribution.

For the Weibull distribution, $T/N \cong M(\Delta T) = \Gamma(1+1/k)/\lambda^{1/k}$ (the value $T/N$ is a statistical estimation of $M(\Delta T)$ and $\lambda_0$), where $T$ is a considered period of seismic events registration and $N$ is the number of events on the scale level. To obtain the G-R law analogue, it is necessary to find the logarithm of the last expression:

$$\log_{10}(N/T) \cong -\log_{10}[M(\Delta T)] = (\log_{10} \lambda)/k - \log_{10}\left[\Gamma(1+1/k)\right]. \qquad (10)$$

Thus to obey the G-R law, it is also enough to have a constant value of the shape parameter, $k$, and a linear relationship between $\log_{10} \lambda$ and energy class.

To test the possibility of the Weibull distribution parameter $k$ being constant for all energy intervals with a maximum-likelihood estimation of $\lambda$, the Kolmogorov test for composite hypotheses has been used (Lemeshko and Postovalov, 2001). According to Table 2, the requisite coincidence between the empirical distribution and the Weibull distribution with constant a $k$ parameter, generally begins at a point on the energy class completeness threshold, $\log_{10} E_{\min}(J)$. The number of strongest events for each region is not sufficient to determine the value of $k$ exactly, and for them it is also possible to have $k<1$ corresponded to an events clustering, like for weak events. Unfortunately, variations in $k$ (Table 2) and $\sigma(\log_{10} \Delta T)$, which is defined by $k$, for different regions are too large, which means that it is not possible to write one unified scaling law for all regions.

A check on the linear relationship between $\log_{10} \lambda$ and energy class (with fixed $k$) was carried out simultaneously with the check on the G-R law. For this purpose, plots of values $-\log_{10}(M(\Delta T)) = (\log_{10} \lambda)/k - \log_{10}[\Gamma(1+1/k)]$ and $\log_{10}(N/T)$ (corresponding to the logarithm of the maximum likelihood estimation for parameter $\lambda_0$ of exponential distribution) were constructed. Both plots should be linear in representative energy intervals (intervals without loss of any events) and should coincide with each other. These properties have been observed in all regions (although the plot for the North Ural Bauxite Mine is bimodal, see above) (Fig. 4). Hence the assumption about the IETSL distribution self-similarity for different scale levels is confirmed, with a scale coefficient $\varphi(E) = d/E^{\gamma} = \lambda^{1/k}$.

To obtain a more explicit analogue of the G-R plot Eq. (8) and first best confirmed property of accelerated life model (7) should be substituted in Eq. (10):

$$\log_{10}(N/T) = -\gamma \log_{10} E - \delta_0 - \log_{10}\left[\Gamma(1+1/k)e^{C/k}\right], \tag{11}$$

where $\gamma$ is a parameter of the accelerated life model, which coincides with the slope coefficient of the G-R plot. Additional investigation has showed that Eq. (10), with a fixed $k$ (see Table 2) and maximum-likelihood estimation for $\lambda$, is a good approximation of the recurrence interval plot.

The expressions obtained for $\log_{10}(N/T)$, written with the parameters of the Weibull distribution (10) and (11) allow the reason for the common deviation from the G-R law in areas of maximum magnitude (Ulomov et al., 1999) to be inferred. Mathematically, such deviation can be related to an increase of the $k$ parameter for the strongest events. This corresponds to changes in the regime of the seismic process in this level, because these events do not prepare stronger events and at this level there is no interaction between events.

Analysis of the Weibull distribution allows the division of points in $\mu_3$ versus $\chi$ plots (see Fig. 2) into two groups with different coefficients of variation to be explained. For the Weibull distribution, $\chi(\Delta T) = \sigma(\Delta T)/M(\Delta T) = [\Gamma(1+2/k) - \Gamma^2(1+1/k)]^{0.5}/\Gamma(1+1/k)$, but for $k = 0.5...1.5$, $\chi(\Delta T) \approx 1/k$. Thus a decrease in the aftereffect for strong events, with a local increase of $k$ (see Fig. 3), leads to a decrease in $\chi(\Delta T)$. This notion links well with general ideas about the coefficient of variation (Kagan and Jackson, 1991). Thus, for weak events, $\chi(\Delta T)$ is >1.3, but for the strongest ones it is ~1.

# 7 Conclusions

The results obtained from the investigation of seismic process over a scale interval covering 14 orders of magnitude of radiated seismic energy, indicate that:

- the seismic process is self-similar in time over a wide range of scales: inter-event times for different scale levels have the same type of distribution, with differences only in scale coefficient, determined by the value of this level (the value of radiated seismic energy); significant deviations occur only in the area of maximum magnitude;

- the distribution of inter-event times for different scale levels is well approximated by the Weibull distribution. At the same time, the exponential distribution approximates well to the strongest events only. This fact and the difference in the value of the coefficient of variation $\chi$ for weak ($\chi > 1$) and strong ($\chi \approx 1$) events gives evidence of the grouping tendency of weak events and the uncorrelated character of the strongest ones;

- the determined relationships allow an equation corresponding to Gutenberg-Richter law to be derived. This confirms the consistency and validity of the results obtained;

- a detailed analysis of inter-event times, with divisions into scale levels as in the G-R law, allows information contained in G-R plot and in the kinetics of inter-event times without division to scale levels to be combined and provides the additional possibility of strong events forecasting.

**Acknowledgements**

The author is very grateful to Prof. V.A. Mansurov for provision of induced seismicity catalogues and Hugh Rice for language corrections.

Table 1. Characteristics of regions considered

| | Induced seismicity | | Natural seismicity: seimoactive regions | | | |
|---|---|---|---|---|---|---|
| Region | North Ural Bauxite Mine | Gold fields Welkom (RSA) | Toktogul | Baikal | Kamchatka | Kamchatka-Kurils |
| Analyzed period, years | 1984…1989 | 01.01.1995…31.08.1995 | 1965…1991 | 1962…1990 | 1962…1990 | 1962…1990 |
| Location | X: 6 400…7 130 m Y: 1 600…1 940 m Z: –110…–980 m | X: 900…19 600 m Y: 350…15 000 m Z: –490…–3 920 m | 39.8…42.7° N.lat. 69.9…74.5° E.lon. | 51…60.6° N.lat. 100…120° E.lon. | 51…58° N.lat. 154…165° E.lon. | Area with apexes, (°N.lat.;° E.lon.): (42; 140), (42; 153), (57; 167), (57; 158), (48; 151), (43; 140) |
| Number of events | 1 200 | 5 400 | 8 800 | 1 700 | 6 100 | 9 500 |
| Energy class of events | 1.6…6.9 | 1.3…10.3 | 1.3…13.9 | 10.3…14.4 | 10.3…18.4 | 10.3…18.8 |
| Energy class completeness threshold $\log_{10} E_{\min}(J)$ | 3.3* | 5.4 | 8.0 | 10.8 | 10.8 | 13.0 |
| Interval obeys accelerated life model | 3.3…5.7 | 5.4…8.5 | 8.0…10.3 | 10.8…14.4 | 10.8…15.2 | 13.0…16.7 |

* here and further the central point of energy interval is written

Table 2. Energy class over which data obey to Weibull, gamma, lognormal and exponential distributions for confidence level more or equal 0.001

| Region | Distribution | | | | | |
|---|---|---|---|---|---|---|
| | $k$ | Weibull | | Gamma | Lognormal | Exponential |
| | | for non fixed $k$ | for fixed $k$ | | | |
| | $S_c$ | 1.2784 | 1.6471 | 1.5651 | 1.6301 | 1.5593 |
| | | Energy class | | | | |
| Induced seismicity | | | | | | |
| North Ural Bauxite Mine | 0.8 | 2.2 | 2.2 | 2.2 | 4.2 | 5.0 |
| Gold fields Welkom (RSA) | 1.0 | 5.8 | 6.0 | 5.8 | 6.2 | 6.0 |
| Natural seismicity: seimoactive regions | | | | | | |
| Toktogul | 0.8 | 7.1 | 7.4 | 9.2 | 9.9 | 12.3 |
| Baikal | 0.8 | 10.3 | 10.3 | 10.3 | 11.8 | 11.2 |
| Kamchatka | 0.65 | 12.6 | 12.7 | 10.3 | 14.9 | 15.0 |
| Kamchatka-Kurils | 0.6 | 13.4 | 13.6 | 13.0 | 16.7 | 16.5 |

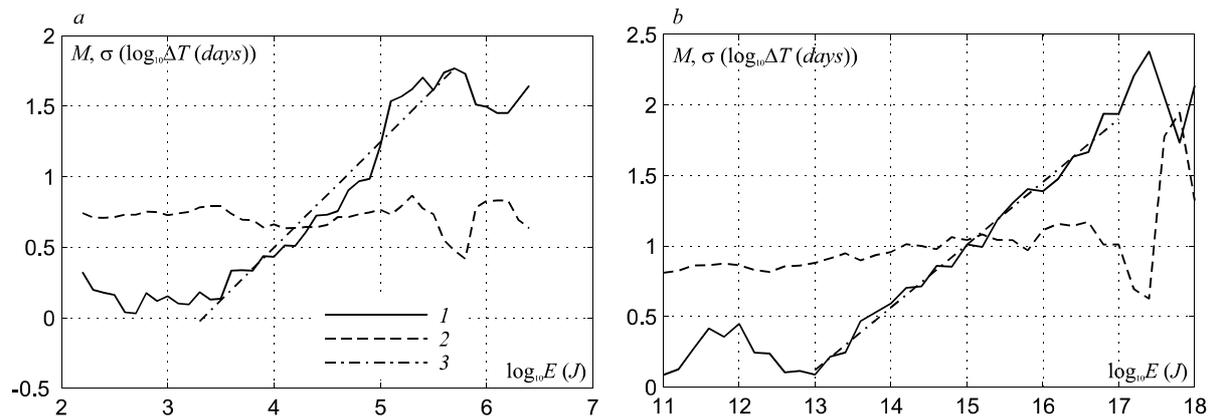

Figure 1. Mean value of logarithm of inter-event times for fixed scale level (1) and standard deviation (2) versus the energy class and a linear least squares approximation for interval, which obey to the accelerated life model (3): *a* – North Ural Bauxite Mine, *b* – Kamchatka-Kurils seismoactive region.

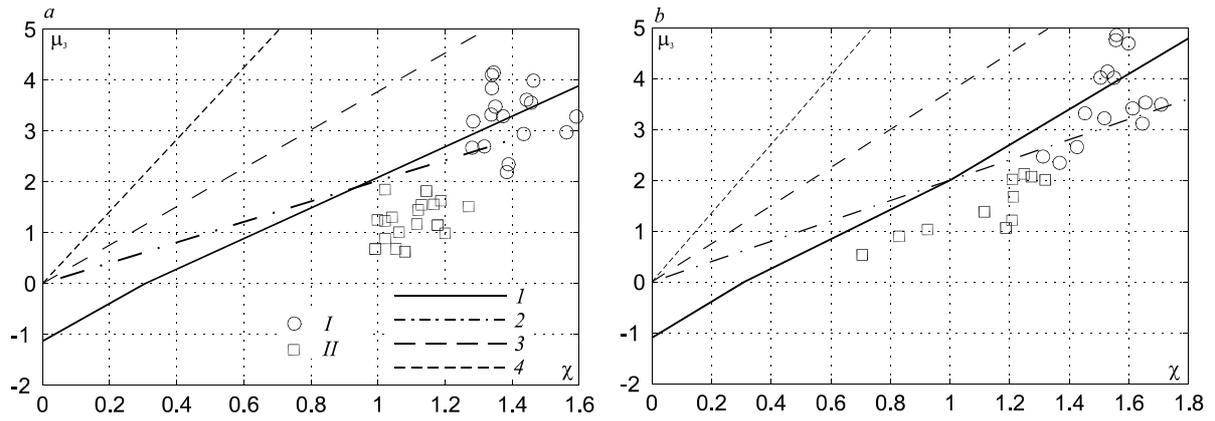

Figure 2. Comparison of the experimental points with theoretical curves (normalized central moment $\mu_3$ versus coefficient of variation $\chi$) for some families of distributions: Weibull (1), gamma (2), lognormal (3), loglogistic (4). A point (1; 2) corresponds to the exponential distribution. *a* – North Ural Bauxite Mine: I – for scale intervals with $\log_{10}E(J) < 5.0$, II – with $\log_{10}E(J) \geq 5.0$; *b* – Kamchatka-Kurils seismoactive region: I – for scale intervals with $\log_{10}E(J) < 16.0$, II – with $\log_{10}E(J) \geq 16.0$.

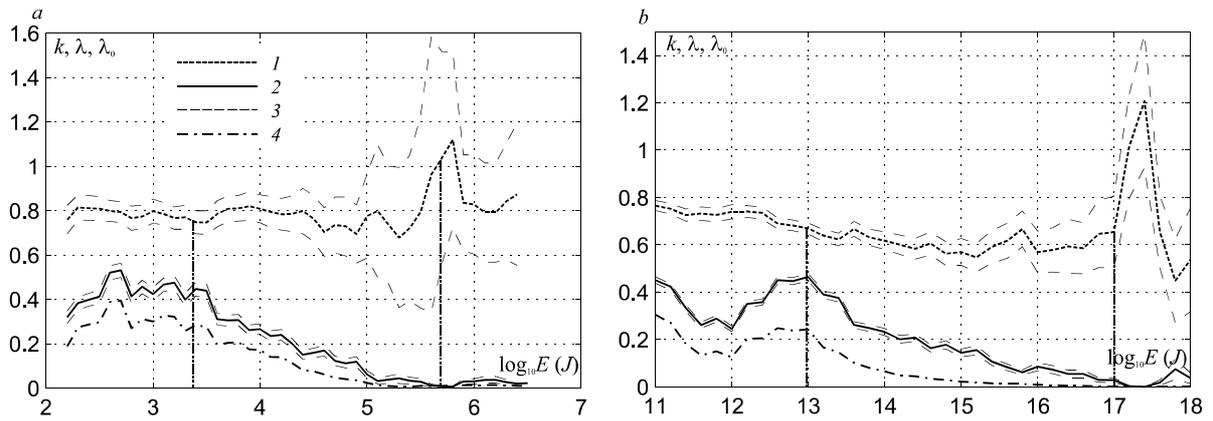

Figure 3. Behavior of the parameters of the Weibull distribution: $k$ (1) and $\lambda$ (2) with corresponded confidence intervals for confidence level 0.1 (3) and the parameter of the exponential distribution $\lambda_0$ (4) versus energy class. Interval which corresponds to the accelerated life model is marked by vertical lines: $a$ – North Ural Bauxite Mine, $b$ – Kamchatka-Kurils seismoactive region.

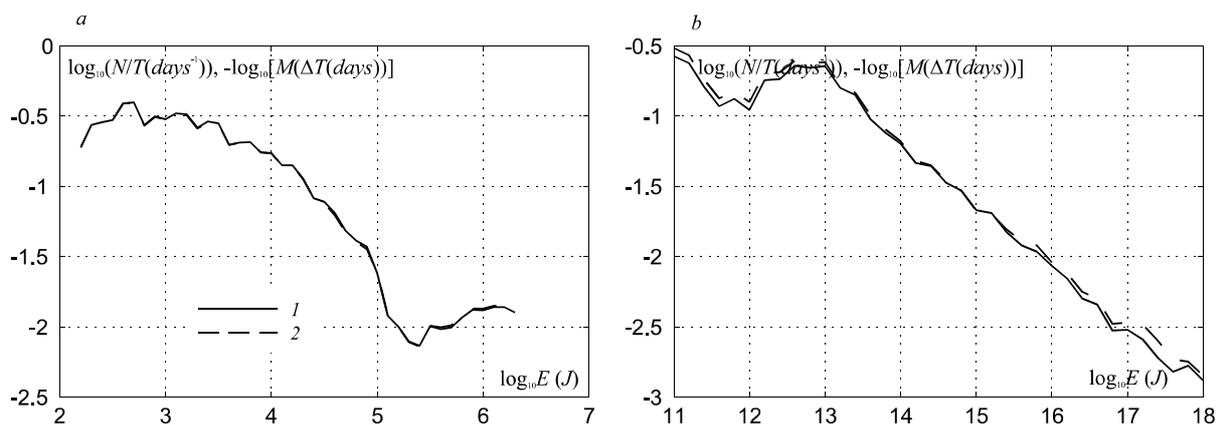

Figure 4. Comparison of the relationship $\log_{10}(N/T)$ corresponded to the G-R plot (1) with the analogue $-\log_{10}(M(\Delta T))$ written through the maximum-likelihood estimation of the scale parameter of the Weibull distribution λ with fixed shape parameter *k* (2), and check of the linear relationship between $\log_{10}\lambda$ and energy class. *a* – North Ural Bauxite Mine, *b* – Kamchatka-Kurils seismoactive region.